\documentclass{emulateapj}
\usepackage{graphicx}
\usepackage{natbib, aas_macros}
\usepackage{multirow}
\usepackage{color}

\usepackage{amsmath}
\def \Vec#1{\mbox{\boldmath $#1$}}

\begin{document}

\title{A search for nontoroidal topological lensing in the Sloan Digital Sky Survey quasar catalog}
\author{Hirokazu Fujii \& Yuzuru Yoshii}
\affil{Institute of Astronomy, University of Tokyo, 2-21-1 Osawa, Mitaka, Tokyo 181-0015, Japan}
\email{hfujii@ioa.s.u-tokyo.ac.jp}
\begin{abstract}
Flat space models with multiply connected topology, which have compact dimensions, are tested against the distribution of high-redshift ($z \geq 4$) quasars of the Sloan Digital Sky Survey (SDSS). 
When the compact dimensions are smaller in size than the observed universe, topological lensing occurs, in which multiple images of single objects (ghost images) are observed. 
We improve on the recently introduced method to identify ghost images by means of four-point statistics. 
Our method is valid for any of the 17 multiply connected flat models, including nontoroial ones that are compactified by screw motions or glide reflection. 
Applying the method to the data revealed one possible case of topological lensing caused by sixth-turn screw motion, however, it is consistent with the simply connected model by this test alone.
Moreover, simulations suggest that we cannot exclude the other space models despite the absence of their signatures. 
This uncertainty mainly originates from the patchy coverage of SDSS in the South Galactic cap, and the situation will be improved by future wide-field spectroscopic surveys. 
\end{abstract}
\shorttitle{Topological Lensing in the SDSS QSO Catalog}
\shortauthors{Fujii \& Yoshii}
\maketitle

\section{Introduction} \label{introduction}

It is widely accepted that the local geometry of the Universe is well described by a homogeneous and isotropic Friedman-Lema\^itre-Robertson-Walker (FLRW) metric with perturbations. 
The concordance $\Lambda$CDM model has achieved remarkable success in explaining various observations such as the faint galaxy number counts \citep[e.g.,][]{1990ApJ...361L...1F,1993ApJ...403..552Y,1995ApJ...444...15Y}, the large-scale distribution of galaxies \citep[e.g.,][]{1990Natur.348..705E,2004ApJ...606..702T,2010MNRAS.401.2148P}, the $m$--$z$ relation for type Ia supernovae \citep[e.g.,][]{1998AJ....116.1009R,1999ApJ...517..565P}, gravitational lensing statistics \citep[e.g.,][]{1999ApJ...510...42C,2012AJ....143..120O}, and the cosmic microwave background (CMB) anisotropies \citep[e.g.,][]{2001PhRvL..86.3475J,2012arXiv1212.5225B}. 
According to the model, our Universe has zero curvature ($\Omega_{\mathrm{tot}}=\Omega_m + \Omega_{\Lambda}=1$) and is described by Euclidean geometry. 
However, the overall shape of space, i.e., the space topology, remains unknown. 
As long as we assume space to be exactly flat and simply connected, i.e., any closed curve on it can continuously shrink to a single point, it inevitably has infinite volume, which is theoretically unfavorable \citep[e.g.,][]{1984NuPhB.239..257H, 1984MNRAS.207P..23Z, 2004JCAP...10..004L}. 
This motivates us to consider multiply connected space models that have compact dimensions \citep[e.g.,][]{1971GReGr...2....7E, 1995PhR...254..135L}. 

One such model is the flat 3-torus $\mathbb{T}^3 = S^1 \times S^1 \times S^1$, which can be visualized as a parallelepiped whose opposite surfaces are formally identified. 
In this space, photons passing through one surface immediately return from the opposite surface, and there are many paths by which light can travel from one point (e.g., a galaxy) to another (e.g., the observer). 
Therefore, if the side length or the compact dimension size is sufficiently small, we would observe multiple images of a single object, which we refer to as {\it ghosts}. 
This phenomenon, often called topological lensing in analogy with gravitational lensing, was first mentioned by Schwarzschild in 1900 \citep[an English translation by][]{1998CQGra..15.2539S}. 
For details on the observation in multiply connected spaces, see an extensive review by \cite{1995PhR...254..135L}. 

Detection of ghost images is the strongest evidence that our space has a compact dimension, and it can be used to determine or constrain the space topology. 
Moreover, it enables us to study the same object at different redshifts, so that we can directly trace the evolution of galaxies. 
Some authors attempted to identify the ghosts of specific objects such as the Milky Way and the rich galaxy clusters \citep{1974JETP...39..196S, 1987MNRAS.224..527D,1987ApJ...322L...5F, 1997MNRAS.292..105R, 2003MNRAS.342L...9W}, while others applied statistical methods to the spatial distributions of cosmic objects \citep{1996A&A...313..339L, 1996MNRAS.283.1147R, 1999A&A...351..766U, 2005A&A...435..427M, 2005JCAP...10..008M}. 
Unfortunately, no robust detections have been reported thus far, which may be reasonable because their methods either (both) need a large number of ghosts or (and) are only valid for toroidal topologies (Section \ref{3Dmanifolds}). 

Over the last decade, the trend in ghost hunting has been to search the CMB map for ghost circles ({\it circles-in-the-sky} method; \citealt{1998CQGra..15.2657C}\footnote{For the historical accuracy, we note that the preprint version of Cornish et al. (1996), arXiv:gr-qc/9602039, described the method for the first time.}), which would arise from the self-intersection of the surface of last scattering along the compact dimension.
 This method and its varieties were applied to the all-sky CMB data obtained by space missions such as the {\it COsmic Background Explorer (COBE)}, the {\it Wilkinson Microwave Anisotropy Probe (WMAP)}, and the {\it Planck Surveyor}.
Although some authors claimed that antipodal ghost circles are likely to be excluded (\citealt{2004PhRvL..92t1302C, 2007PhRvD..75h4034K, 2011MNRAS.412.2104B}; \citealt{2013arXiv1303.5086P}), diverse results have also been obtained \citep{2008CQGra..25v5017A, 2008A&A...486...55R}. 
\cite{2012PhRvD..86h3526V} extended the search to general (including nonantipodal circles) cases for the first time. 
 Their analysis, however, is based on the coarse-grained ($N_{\mathrm{side}}=128$ in the HEALPix scheme; \citealt{2005ApJ...622..759G}) map, and it is not clear whether the topological signatures are detectable in such low-resolution maps. 

Based on these situations, we consider it important to subject the multiply connected space models to different tests, especially for the nontoroidal models (see Section \ref{3Dmanifolds}), which have been studied very little in past works.  
This paper presents a new, object-based method that is valid for these models. 
It is an extension of the works of \cite{2011A&A...529A.121F} and \cite{2012A&A...540A..29F}, and herein, we apply it to a high-redshift ($z \geq 4$) quasar sample of the Sloan Digital Sky Survey (SDSS) data release 7 (DR7; \citealt{2010AJ....139.2360S}). 
We consider this work timely because massive spectroscopic surveys such as SDSS have become increasingly important in recent cosmological studies. 

All calculations herein are done considering comoving space. 
The cosmological parameters of the concordance model, $\Omega_m = 0.27$, $\Omega_{\Lambda}=0.73$, and $h=0.71$, are used when translating redshift into comoving distance.  
An incorrect selection of values of $(\Omega_m, \Omega_{\Lambda}, h)$ distorts the comoving distribution of objects; thus, we generally have to perform our analyses for a number of cosmological models. 
However, we checked that our results do not change within the errors (a few percent) of the concordance model. 
Unless otherwise stated, to represent a position in space we use the Cartesian coordinates 
\begin{equation} \label{Coordinates}
x = r \cos \delta \cos \alpha, \ y = r \cos \delta \sin \alpha, \ z = r \sin \delta,
\end{equation}
where $r, \alpha$, and $\delta$ are the comoving distance, the right ascension (RA), and the declination (Dec), respectively. 

A brief summary of three-dimensional flat manifolds is given in Section \ref{3Dmanifolds}. 
In Section \ref{method}, we describe the method for detecting topological lensing effects in flat spaces. 
The descriptions of our sample and simulated catalogs are given in Section \ref{results}, followed by the results and discussion in Section \ref{discussion_and_conclusion}.

\section{Three-dimensional flat manifolds} \label{3Dmanifolds}
Since we assume the concordance model with zero curvature, it is helpful to summarize the mathematical classification of three-dimensional (3D) flat manifolds. 
It has been proved that there are a total of 18 space forms \citep{nowacki1934euklidischen}, including the simply connected Euclidean space $\mathbb{E}^3$. 
Figure 1 of \cite{2011A&A...529A.121F}, which was first published in \cite{cipra2002whats}, displays an illustration of the 17 multiply connected space forms, where the surfaces are identified in pairs by Euclidean isometries. 

Mathematically, they are the quotients $\mathbb{E}^3/\Gamma$ of the simply connected space $\mathbb{E}^3$, where $\Gamma$ is a group of Euclidean isometries that are discrete and fixed-point free \citep[e.g.,][]{9783540152811}. 

\begin{figure}
\centering
\includegraphics[bb=50 85 550 605, width=7.5cm]{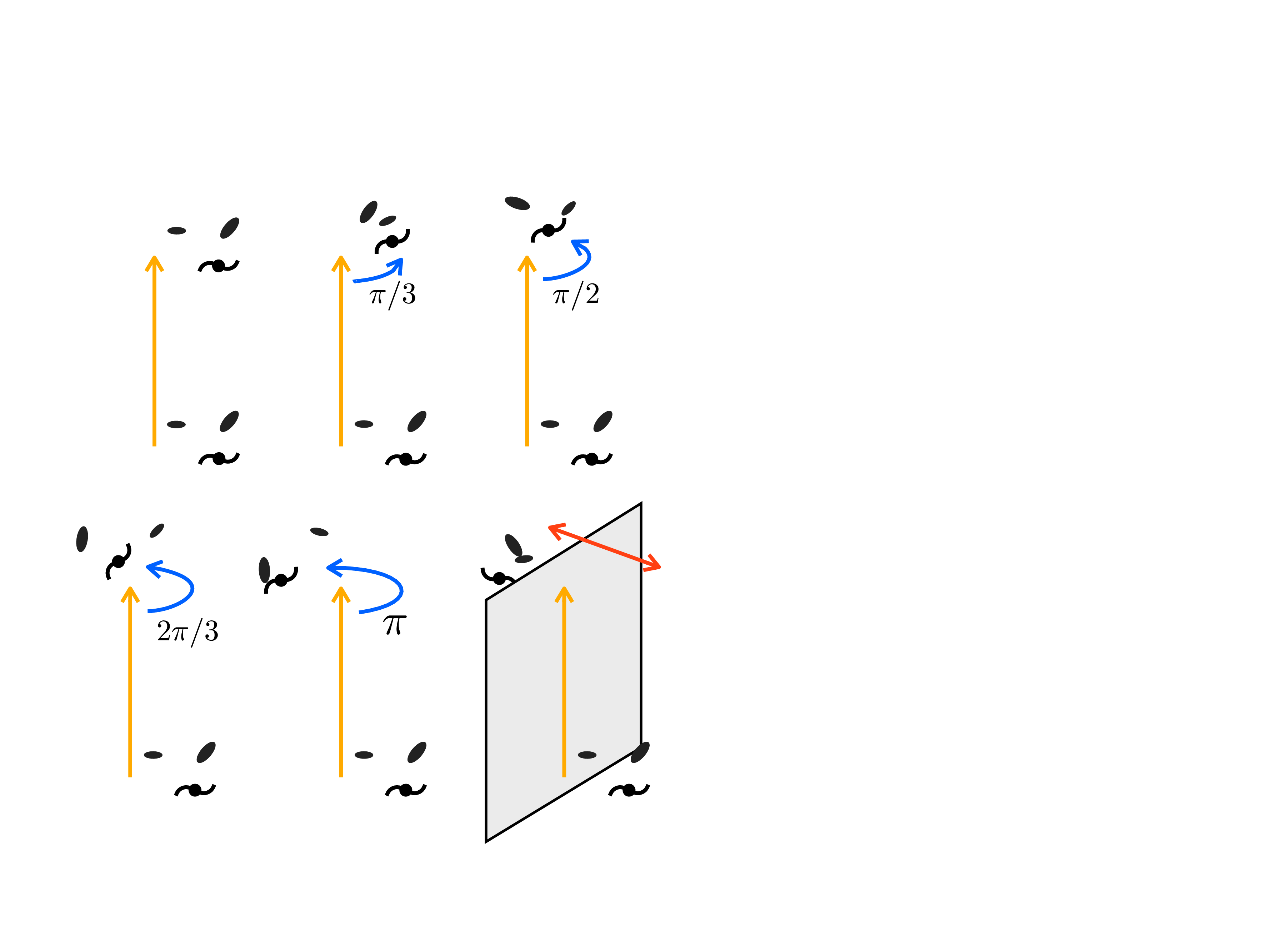}
\caption{The fundamental isometries of 3D flat manifolds: translation, screw motions of twists $\theta_{\mathrm{twist}}=\pi, 2 \pi/3, \pi/2$, and $\pi/3$, and glide reflection. From these isometries, we can generate 17 nontrivial groups that are discrete and fixed-point free; these correspond to the 17 multiply connected space models.} \label{Isometries}
\end{figure}

The isometries consist of three types: translation, screw motion (rotation followed by translation), and glide reflection (reflection followed by translation). 
Because of symmetry, the possible twists of the screw motion are $\pi$, $2\pi/3$, $\pi/2$, and $\pi/3$. 
Therefore, we need only consider six kinds of isometries (Figure \ref{Isometries}) to put constraints on the 17 multiply connected models\footnote{Strictly speaking, some manifolds are generated by other types of isometries, e.g., the Hantszche-Wendt space is generated by half-turn screw motions whose screw axis is not parallel to the translation direction. We do not consider such varieties separately, since they can be detected by our method as one of the six isometries displayed in Figure \ref{Isometries}.}. 
Although the right-handed screw motions are displayed in Figure \ref{Isometries}, they might be actually observed as left-handed. The two cases are intrinsically the same, and the handedness distinction totally depends on our notion of right and left (Jeffrey Weeks 2013, private communication). 

Ghost images are linked to each other by the isometries as seen in Figure \ref{Isometries}, which would leave a specific pattern in the distribution of objects. 
The toroidal models, whose isometries are all translations, are relatively easy to constrain, since ghost images form a simple lattice pattern and ghost circles are always antipodal. 
Indeed, most past works limited their targets to these models and there are some useful methods available (e.g., \citealt{2005A&A...435..427M, 2005JCAP...10..008M}). 
The isometries of nontoroidal models, in contrast, include screw motions and glide reflections, and the pattern of topological lensing depends on the observer's location relative to the screw axis or the reflection plane. 
In this work, we explore the possibility that our space has a nontoroidal topology using high-redshift quasars. 
Although the space models compactified by glide reflection are nonorientable and physically unreasonable \citep[e.g., ][]{demianski2004topology}, we do not exclude these cases from our targets. 

\section{The four-point statistics method for detecting topological lensing} \label{method}

In this section, we propose a method to find ghost images from 3D positions of objects, in which a kind of four-point statistical analysis is performed. 
It is essentially based on the works of \cite{2011A&A...529A.121F} and \cite{2012A&A...540A..29F}, however, some modifications are added that enhance the signal-to-noise ratio. 
By using this method, any of the Euclidean isometries is detectable with high sensitivity.  

Suppose that we have a sample of $N$ discrete sources whose comoving positions are $\Vec x_1, \Vec x_2, \cdots$, and $\Vec x_N \in \mathbb{E}^3$ in the Cartesian coordinates given by Equation (\ref{Coordinates}). 
The separations between all pairs of points are calculated to identify {\it isometric twins}, i.e., a pair of twins $(\Vec x_i, \Vec x_j)$ and $(\Vec x_k, \Vec x_l)$ that have (nearly) equal distance: $| \Vec x_i - \Vec x_j | \approx | \Vec x_k - \Vec x_l |$. 
They are possibly the topologically lensed images of each other, or {\it ghost twins}, since any isometry preserves distance so that ghost twins are always isometric. 

When the Universe is multiply connected and we have a large number of ghosts, there would be more isometric twins (isometric $n$-tuples in general; see \citealt{1996MNRAS.283.1147R}) than normally expected. 
This could be used as a test for multiply connected space models \citep{1999A&A...351..766U}, but in practice it is hard to detect the slight difference. 
For example, the number of isometric twins in our SDSS sample (Section \ref{results}), within the tolerance of $10 \ \mathrm{Mpc}$, is as large as $\approx 2 \times 10^9$. 
It is almost impossible to distinguish simply connected from  multiply connected space models by the number of isometric twins, unless the Universe is so small that the topological lensing effect is very strong, i.e., a large number of ghosts corresponding to a single object can be seen. 

This difficulty, however, can be overcome by classifying the isometric twins according to their {\it linking isometries}. 
Given a pair of isometric twins $(\Vec x_i, \Vec x_j)$ and $(\Vec x_k,\Vec x_l)$, an isometry $\gamma$ is called the linking isometry  when $\gamma$ links the two, i.e., $(\Vec x_k, \Vec x_l) = (\gamma \Vec x_i, \gamma \Vec x_j)$. 
If the space is multiply connected and compactified by an isometry $\gamma$, the number of isometric twins that are linked by $\gamma$ would be higher than normally expected; this is the basic idea of our method. 
Calculation of linking isometries proceeds as follows.  


\paragraph{1. Translation}
When the twins are equal as vectors, i.e., $\Vec x_j - \Vec x_i \approx \Vec x_l - \Vec x_k,$ they can be linked by a translation. 
The translation is fully characterized by the vector 
\begin{equation}
\Vec L = L \Vec n = \frac{(\Vec x_k - \Vec x_i)+(\Vec x_l - \Vec x_j)}{2},
\end{equation}
where $\Vec n$ denotes the direction of translation and $L= | \Vec L |$ corresponds to the compact dimension size. 
The parameter space for translation, $(\Vec n, L)$, is three dimensional. 
We do not consider the translational isometries hereafter. 

\paragraph{2. Screw motion}

When the twins make an angle of 
\begin{equation}
\theta_{\mathrm{int}}=\cos ^{-1} \biggl[ \frac{(\Vec x_j - \Vec x_i) \cdot (\Vec x_l - \Vec x_k)}{| \Vec x_j - \Vec x_i | | \Vec x_l - \Vec x_k |} \biggr],
\end{equation}
they can be linked by a screw motion of twist $\theta_{\mathrm{twist}} \geq \theta_{\mathrm{int}}$, where the meaningful cases are $\theta_{\mathrm{twist}} = \pi, 2\pi/3, \pi /2$, and $\pi/3$. 
A screw motion is characterized by a screw axis and a translation length. 

The direction of the screw axis, $\Vec n$, is calculated first. 
This is obtained by rotating a unit vector parallel to $(\Vec x_j - \Vec x_i)+(\Vec x_l -\Vec x_k)$ about the axis parallel to $(\Vec x_j - \Vec x_i)-(\Vec x_l -\Vec x_k)$ by an angle of $ \sin ^{-1} [ \tan (\theta_{\mathrm{int}} /2)/ \tan (\theta_{\mathrm{twist}} /2 ) ]$ or its supplementary angle (so there are two cases). 
Seen from this direction, the two twins apparently make an angle of $\theta_{\mathrm{twist}}$. 
The screw axis is represented by the equations 
\begin{equation}
\Vec x = \frac{\Vec F (\Vec x_i, \Vec x_k ) + \Vec F(\Vec x_j, \Vec x_l)}{2} + t \Vec n, \label{AxisEquation}
\end{equation}
where $t \in ( - \infty , + \infty) $ is a parameter and $\Vec F (\Vec x, \Vec y)$ is a vector function defined by
\begin{eqnarray}
\Vec F(\Vec x, \Vec y) &\equiv& \frac{\Vec x + \Vec y}{2} \nonumber \\
&+& \frac{| \Vec x - \Vec y - [ ( \Vec x - \Vec y ) \cdot \Vec n ] \Vec n |}{2 \tan (\theta _{\mathrm{twist}} /2 )} \frac{(\Vec x - \Vec y) \times \Vec n}{|(\Vec x - \Vec y) \times \Vec n|}.
\end{eqnarray}
We then introduce a new coordinate system $(X, Y, Z)$ whose basis vectors are $\Vec e_Z = \Vec n, \ \Vec e_Y = (\Vec e_Z \times \Vec e_z)/|\Vec e_Z \times \Vec e_z| $, and $\Vec e_X = \Vec e_Y \times \Vec e_Z$, and whose origin is the Milky Way. 
All points on the screw axis have the same $(X, Y)$ coordinates, which give the position of the axis. 
Finally, the translation length is given by 
\begin{equation}
L = \frac{[(\Vec x_k - \Vec x_i)\cdot \Vec n] + [(\Vec x_l - \Vec x_j) \cdot \Vec n]}{2},
\end{equation}
which corresponds to the compact dimension size. 
In this way, given a pair of isometric twins, the linking screw motion of twist $\theta_{\mathrm{twist}}$ is determined. The parameter space of screw motion, $(\Vec n, X, Y, L)$, is five dimensional. 


\paragraph{3. Glide reflection}

Any pair of isometric twins can be linked by a glide reflection. 
A glide reflection is characterized by a reflectional plane and a translation vector on it. The normal vector of the plane, $\Vec n$, is given by the unit vector parallel to $(\Vec x_j - \Vec x_i)-(\Vec x_l -\Vec x_k)$. 
By introducing the same $(X, Y, Z)$ coordinates as in the case for screw motion, all points on the reflectional plane have the same $Z$ coordinate:
\begin{equation}
Z = \frac{ (\Vec x_i + \Vec x_j + \Vec x_k + \Vec x_l )\cdot \Vec n}{4},
\end{equation}
which gives the position of the plane. Finally, the translation vector is given by
\begin{eqnarray}
\Vec L &=& \frac{(\Vec x_k - \Vec x_i) - [(\Vec x_k - \Vec x_i)\cdot \Vec n]\Vec n}{2} \nonumber \\
&+&\frac{(\Vec x_l - \Vec x_j) - [(\Vec x_l - \Vec x_j)\cdot \Vec n]\Vec n}{2}. 
\end{eqnarray}
Since this vector always satisfies $\Vec L \cdot \Vec n = 0$, it has only two degrees of freedom: $L_X$ and $ L_Y$. 
The glide reflection that links the twins is determined in this way. 
The parameter space of glide reflection, $(\Vec n, Z, L_X, L_Y)$, is again five dimensional.

\

After calculating the linking isometries for all pairs of isometric twins, we then search for excess counts in the parameter space for each type of isometry. 
We describe the process for the quarter-turn screw motion ($\theta_{\mathrm{twist}}= \pi/2$) here; the other cases can be studied in the same way. 
First, for each object $\Vec x_i$, all pairs of isometric twins that include $\Vec x_i$ as a member are selected. 
They give a set of points $(\Vec n_1, X_1, Y_1, L_1), \cdots, (\Vec n_{\alpha}, X_{\alpha}, Y_{\alpha}, L_{\alpha})$ in the parameter space of quarter-turn screw motion. 
The five-dimensional (5D) parameter space is binned, and the number of points in a bin at $(\Vec n, X, Y, L)$ is denoted as $s_i (\Vec n, X, Y, L)$. 
To extract bins that have excess counts, we define the ``score'' for each bin as 
\begin{equation}  \label{StepFunc}
\chi (s_i)= \left \{ 
\begin{array}{l}
s_i, \ \mathrm{if} \ s_i \geq s_{\mathrm{min}} \\
0, \ \mathrm{otherwise}.
\end{array}
\right.
\end{equation} 
In this paper, the threshold value was chosen to be $s_{\mathrm{min}}=3$, by which we trade off the detectability of ghost triples for reducing the noise level. 
It is a reasonable choice because isometric triples are too common to be taken as evidence for topological lensing.

The total score is given by
\begin{equation}
S(\Vec n, X, Y, L) = \sum_{i=1} ^N \chi[s_i(\Vec n, X, Y, L)].
\end{equation}	
Finally, we search for a bin whose score is much higher than the stochastic expectation, which might be a sign of multiply connected topology. 
If there is a pair of isometric $n$-tuples (with $n$ higher than $s_{\mathrm{min}}=3$) that is linked by a quarter-turn screw motion, the corresponding score would be $S=2n(n-1)$.

We note that the two-step process, in which we first search for excess counts in $N$ subsets and next sum them, is indeed unessential but inherited from the previous works \citep{2011A&A...529A.121F, 2012A&A...540A..29F}. 
It was originally adopted to favor a pair of isometric $n$-tuples over a large number of isometric twins, because the previous methods could not distinguish the two since the parameters of isometries were not fully used. 
As long as we use the current improved method, the results would not change unless we adopt the two-step process.

\section{Quasar catalogs} \label{results}

\subsection{Observation}
Quasars are one of the most luminous ($M_B \leq -23$) sources in the Universe and can be observed at great distances, which makes them a good probe to constrain the global structure, i.e., the topology, of space. 
Based on the four-point statistics method (Section \ref{method}), we aim to search for ghost images in the quasar catalog of the SDSS DR7 \citep{2010AJ....139.2360S}. 
The SDSS  catalog is thus far the most massive spectroscopic set of quasar data, with $\approx 100,000$ objects at redshifts $z \lesssim 5.5$. 
The typical redshift error is $\Delta z \approx 0.004$ \citep{2010AJ....139.2360S}.

\begin{figure}
\centering
\includegraphics[trim=10 10 5 5, width=9.5cm]{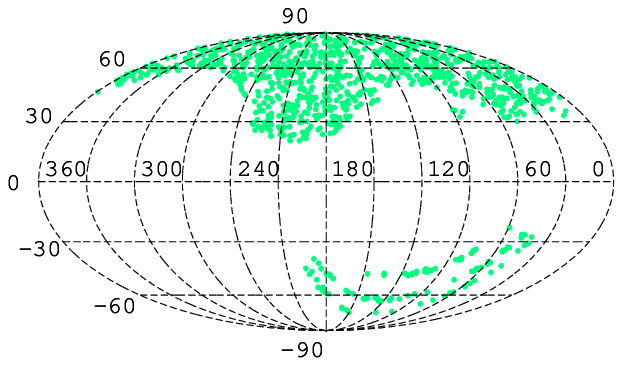}
\includegraphics[trim=10 10 5 5, width=9.5cm]{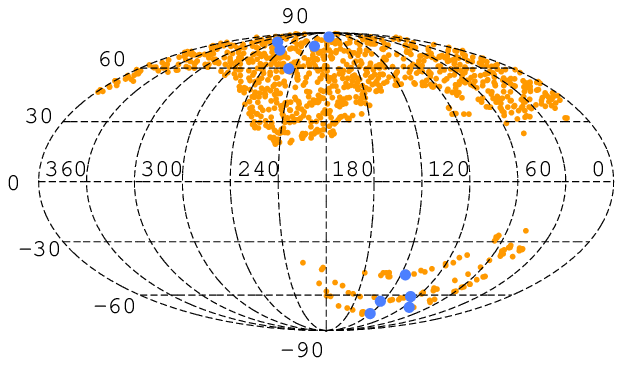}
\caption{The celestial distribution of quasars in the Mollweide projection with Galactic coordinates. The dashed lines represent intervals of $30^{\circ}$. {\it Top:} Our sample selected from the SDSS catalog. {\it Bottom:} One of the mock catalogs that contains a pair of ghost quintuples (blue points). The ghosts are linked by a quarter-turn screw motion whose translation length is $14 \ \mathrm{Gpc}$.}
\label{SkyMap}
\end{figure}

From the original data, the most luminous ($M_i \lesssim -27.7$) 1,000 objects at redshifts $z \geq 4$ were selected, which made the sample almost complete (see the descriptions of imaging and photometric completeness in \cite{2005AJ....129.2047V} and \cite{2006AJ....131.2766R}, respectively) and enables us to explore the compact dimension size up to $15 \ \mathrm{Gpc}$. 
As shown in Figure \ref{SkyMap}, the majority ($\approx \negthickspace90$\%) of items in the sample are in the North Galactic cap, whereas the South Galactic cap is only covered by three narrow stripes. The sky coverage significantly limits our test as described in Section \ref{discussion_and_conclusion}. 
The comoving positions of objects were calculated using the set of cosmological parameters given in Section \ref{introduction}. 

\subsection{Simulations} \label{simulations}
For comparison with the data, mock catalogs in simply connected and multiply connected spaces were generated. 
Since we assume a flat cosmology, the simply  connected space is $\mathbb{E}^3$. 
As an example of multiply connected models, the quarter-turn space, $\mathbb{T}^3/\mathbb{Z}_4$, generated by a quarter-turn ($\theta_{\mathrm{twist}}=\pi /2$) screw motion with $L=14 \ \mathrm{Gpc}$ was chosen (sizes of the other two dimensions that are compactified by translation were set to be larger than the horizon scale). 
The orientation and the position of the screw axis were set to $\Vec n=(-0.953, 0.219, 0.211)$, which corresponds to $(\alpha, \delta)=(167^{\circ}, 12^{\circ})$, and $(X, Y)=(-0.75, 3.25)$ Gpc, respectively, so that the ghost images were located within the sky coverage of SDSS (Figure \ref{SkyMap}). 
Hereafter, all mentions of mock catalogs in multiply connected space refer to this model.  

\begin{figure}
\centering
\includegraphics[trim=0 0 0 0, width=8cm]{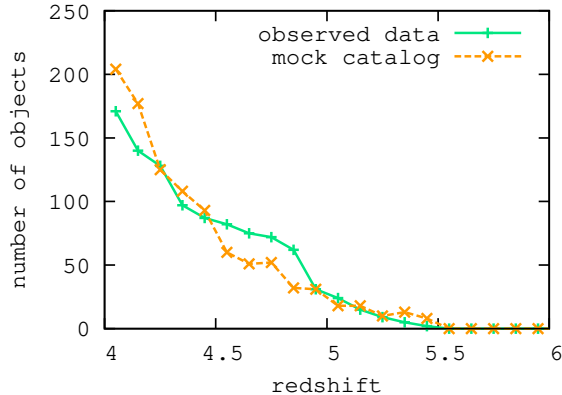}
\caption{The number of quasars in each redshift bin ($\Delta z = 0.1$). {\it Green line:} our sample selected from the SDSS catalog. {\it Orange line:} one of the mock catalogs whose celestial distribution is shown in Figure \ref{SkyMap}. Mock quasars basically follow the same redshift distribution as that of the actual quasars.}
\label{RedshiftDistribution}
\end{figure}

The physical properties of quasars are modeled as follows. 
First, the simple ``light bulb" model was used for the quasar light curve, in which each object emits radiation for a fixed duration $t_{\mathrm{life}} =100 \ \mathrm{Myr}$ with constant luminosity (the {\it i} band absolute magnitude $M_i$). 
To take into account obscuration of the active nucleus by the dust torus, which reduces the number of observable ghosts, each object is covered by a spherical mask with antipodal holes. 
The orientation of the mask is randomly selected, and the opening angle of the holes is given by
\begin{equation}
\theta_{\mathrm{open}} = \cos^{-1} \biggl [ \frac{1}{\sqrt{1+3(L_{5007}/L_0)^{1-2 \xi} }} \biggr ], 
\end{equation}
where $L_{5007}$ is the [OIII] luminosity at wavelength $\lambda = 5007 \ \mathrm{\AA}$, $L_0 = 10^{42.37} \ \mathrm{erg \ sec^{-1}}$, and $\xi = 0.23$ \citep{2005MNRAS.360..565S}. That is, more luminous quasars are more likely to be observed as type 1. 
To convert $M_i$ into $L_{5007}$, the empirical relation implied by the value-added catalog of \cite{2011ApJS..194...45S} was used: 
\begin{equation}
\log L_{5007} = -0.307 M_i + 34.8, 
\end{equation}
but we note that this relation has a large scatter. 

We describe below how to construct mock catalogs, but for simplicity we first consider the simply connected space.
In this case, the mock quasars are randomly\footnote{Although the real quasars are clustered similar to ordinary galaxies \citep[e.g.,][]{2005MNRAS.356..415C, 2007AJ....133.2222S}, the effect of clustering on this test can be canceled by the third condition for selecting four points (Section \ref{discussion_and_conclusion}).} 
generated in the infinite Euclidean space $\mathbb{E}^3$ every 10 Myr, whereas those that reach the end of their lifetime ($=100$ Myr) are removed. 
Each object is assigned the luminosity $M_i$ so that the number density of type 1 quasars follows a luminosity function (LF) of the form                                                                         
\begin{equation} 
\Phi (M_i | z) = \Phi^{*} 10^{A_1 [M_i - (M^{*} + B_1 \xi + B_2 \xi ^2 + B_3 \xi ^3 )]}, 
\end{equation}
where 
\begin{equation}
\xi = \log \biggl( \frac{1+z}{1+z_{\mathrm{ref}} } \biggr ).
\end{equation}
We adopted the maximum likely set of parameters obtained by fitting the LF to the SDSS DR3 quasars \citep{2006AJ....131.2766R}, $A_1 =0.78, B_1=0.1, B_2=27.35, B_3 = 19.27$, and $\log \Phi^{*} = -5.75$ (where $M^{*}$ and $z_{\mathrm{ref}}$ were set to $-26$ and $2.45$, respectively). 
Figure \ref{RedshiftDistribution} shows that the observed redshift distribution is correctly reproduced in our simulation. 

For the multiply connected case, we must also consider the topological lensing effect. It is mimicked by tessellating the infinite space $\mathbb{E}^3$ with identical copies of a polyhedron that represents the whole physical space, i.e., contains all objects but no ghost images, such that the copies are mapped onto each other by the corresponding group of isometries (see Section \ref{3Dmanifolds}). Thus the procedure is as follows: we first distribute mock quasars in one polyhedron in the same way as the simply connected case, and then tessellate $\mathbb{E}^3$ with their copies (i.e., ghosts) so as to match the topology considered.

The positional uncertainty primarily originates from the motion of objects with respect to the Hubble flow, which creates extra redshifts in addition to those of cosmological origin (known as the redshift space distortions).  
We assign each object the peculiar velocity of $v=500 \ \mathrm{km \ sec^{-1}}$, which is a typical value, and with a randomly selected direction\footnote{A Gaussian distribution of peculiar velocity is probably more realistic, but it would not change the ``typical" displacement of comoving positions.}. 
Then, the measurement errors were added to the true redshifts (Hubble flow plus peculiar motion), which are randomly selected from a Gaussian distribution of the rms of $\Delta z=0.004$. 
Because of these effects, the comoving distance deduced from the redshifts deviates from the real value by $\lesssim 10 \ \mathrm{Mpc}$. 

We note that, since ghost images are generally observed at different look-back times, there is another source of the positional uncertainty related to the peculiar velocity: the movement of objects in real space (e.g., \citealt{1996MNRAS.283.1147R,1999A&A...351..766U, 2000A&A...363....1L}).
However, this effect is much weaker (a few percent) than those considered above, since we consider ghost images at nearly the same look-back times (Section \ref{discussion_and_conclusion}).

\section{Results and Discussion} \label{discussion_and_conclusion}

The method described in Section \ref{method} is applied to the distribution of real quasars in this section.
All the pairs of twins, $(\Vec x_i, \Vec x_j)$ and $(\Vec x_k, \Vec x_l)$, that satisfy the following conditions were selected:
\begin{eqnarray}
\ast & \ & | |\Vec x_i -\Vec x_j| - |\Vec x_k - \Vec x_l | | \leq 10 \ \mathrm{Mpc}, \\[0.2cm]
\ast & \ & |t_i - t_k|, |t_j - t_l| \leq 100 \ \mathrm{Myr}, \\[0.2cm] 
\ast & \ & \mathrm{and} \ |\Vec x_i - \Vec x_j|, |\Vec x_k - \Vec x_l| \geq 100 \ \mathrm{Mpc},
\end{eqnarray}
where $t_i, t_j, \cdots$ are the cosmic times corresponding to the comoving positions $\Vec x_i, \Vec x_j, \cdots$, respectively. 
The latter two conditions were not mentioned in Section \ref{method}. 

The first condition selects pairs of isometric twins; ghost twins are always isometric \citep[e.g.,][see also Section \ref{method}]{1996MNRAS.283.1147R, 1999A&A...351..766U} within the tolerance of $\approx 10 \ \mathrm{Mpc}$. 
The second condition is useful when considering short-lived objects such as quasars, since the temporal separations between ghost images cannot be longer than the object's lifetime.
The last one reduces the false positives resulting from clustering: In a strongly clustered region, the mean separation between objects would be so small that any pair of twins would satisfy the first condition.
It would not reduce the topological signature since the real ghosts are unlikely to be clustered in a small region.
Those that are similar to the second and third conditions were also used by \cite{2005A&A...435..427M}, who explored the toroidal topologies using active galactic nuclei (AGNs).
When applied to mock catalogs in multiply connected space, the selection completeness of the above criteria were found to be sufficiently high: More than $99\%$ of ghost twins were correctly selected. 
However, the selection efficiency is as low as $10^{-6}$ and we must classify the selected pairs of twins as described in Section \ref{method}.

\begin{figure}
\includegraphics[trim=0 0 0 5,width=8.5cm]{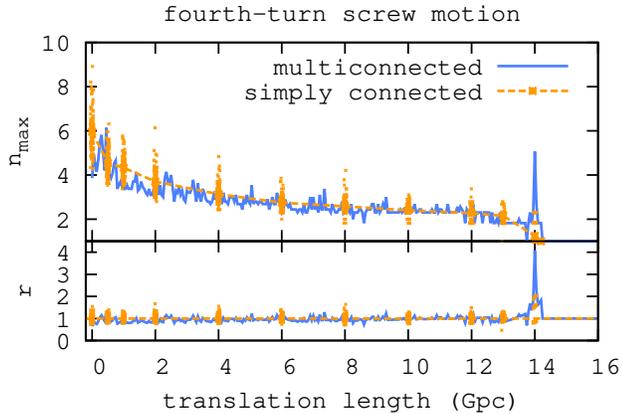}
\caption{
The result of a ghost search for the quarter-turn screw motion. 
\textit{Orange line:} The mean curve over 100 simulations in the simply connected space along with each realization.
\textit{Blue line:} The result for a mock catalog in the quarter-turn space, which contains a pair of ghost quintuples. 
The ghosts are clearly detected as a spike of height $5$ at $L=14 \ \mathrm{Gpc}$. 
The bottom figure is the same as the top one but normalized to the mean result of the simply connected model.  
For the definition of $n_{\mathrm{max}}$, see Equation (\ref{Vertical}) and the text.  
}
\label{ResultMock}
\end{figure}

First, to see how our method works, we applied the method to a mock catalog that contains a pair of ghost quintuples. 
(The celestial distribution of the ghosts is displayed in Figure \ref{SkyMap}.)
We searched for the signature of the quarter-turn screw motion since the ghosts were modeled to be linked by an isometry of this type (see Section \ref{simulations}). 
The result is shown in Figure \ref{ResultMock}, which is plotted with 
\begin{equation}
n_{\mathrm{max}}(L) \equiv \frac{1}{2} + \sqrt{\frac{1}{2} \biggl( S_{\mathrm{max}}(L) + \frac{1}{2} \biggr )} \label{Vertical}
\end{equation}
as the vertical axis, where 
\begin{equation}
S_{\mathrm{max}}(L) = \mathrm{max} \{ S(\Vec n, X, Y, L') \} _{L'=L},
\end{equation}
and with the translation length $L$ as the horizontal axis. 
(Note that a pair of isometric $n$-tuples corresponds to $S=2n(n-1)$ when $n \geq s_{\mathrm{min}}+1=4$.)
For comparison, we also plotted the mean values averaged over 100 realizations in the simply connected space, as well as the ratio between the two curves. 
Each realization is also plotted as small dots to show the variances in the simply connected case.

\begin{figure*}
\centering
\includegraphics[trim= 0 0 0 0, width=8.5cm]{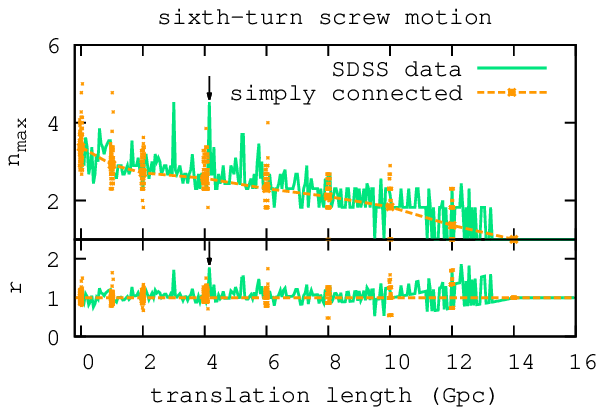}
\includegraphics[trim= 0 0 0 0, width=8.5cm]{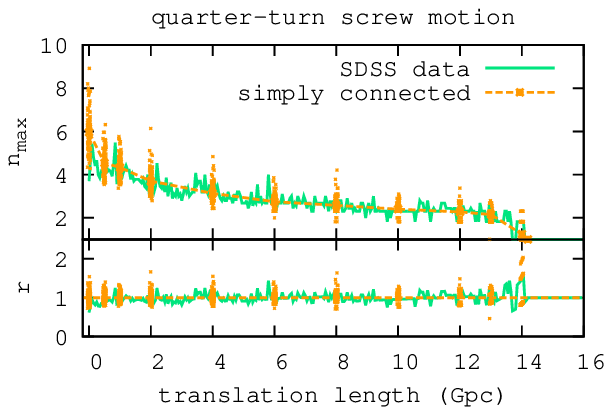}
\includegraphics[trim= 0 0 0 0, width=8.5cm]{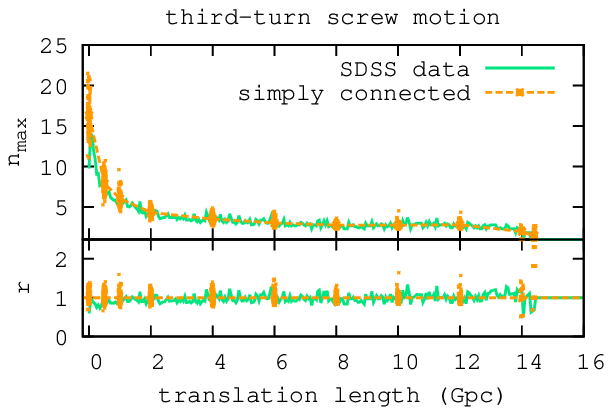}
\includegraphics[trim= 0 0 0 0, width=8.5cm]{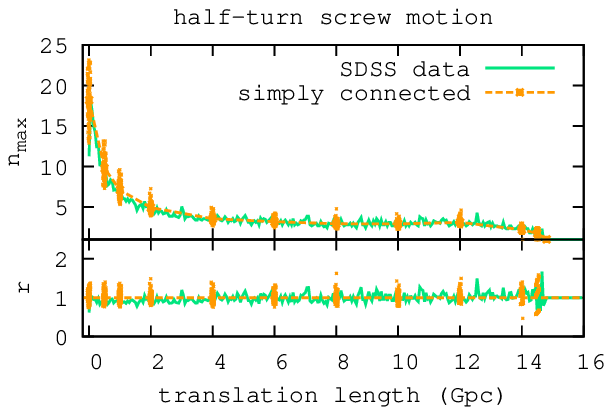}
\includegraphics[trim= 0 0 0 0, width=8.5cm]{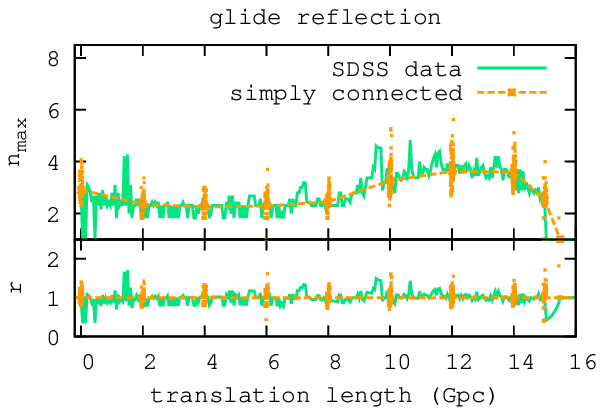}
\caption{Results of the quest for the topological signatures in the SDSS high-redshift ($z\geq 4$) quasar sample. 
Five types of nontranslational Euclidean isometries were explored. 
\textit{Green lines:} The results for the observed data. 
\textit{Orange lines:} The mean curves over 100 simulations in the simply connected space along with each realization plotted as small dots.
In each case, the bottom figure displays the ratio between the observation and the mean result for the simply connected model. 
From these curves we identified one candidate for the topological signature that satisfy $n_{\mathrm{max}}\geq 4$ and $r \geq 1.5$, for the sixth-turn screw motion (marked with vertical arrows; see Table \ref{CandidatesTable} for details). 
However, this is consistent with the simply connected model and we do not detect any ``robust'' signatures such as that seen in Figure \ref{ResultMock}. 
}
\label{ResultData}
\end{figure*}

As expected, we find a sharp spike of height $n_{\mathrm{max}}=5$ at $L=14 \ \mathrm{Gpc}$ that is produced by the ghost quintuples. 
The spike occurs in the 5D parameter space at $\Vec n=(-0.95,0.23,0.23)$, $(X,Y)=(-0.72,3.28) \ \mathrm{Gpc}$, and $L=14 \ \mathrm{Gpc}$, which correctly reproduces the input parameters (Section \ref{simulations}).
This result shows that our method is practically useful for detecting the signature of nontoroidal topologies, without any assumptions on the orientation of the Universe or on the location of the observer.  
If the observed catalog contains some number of ghost images, we would surely be able to identify them. 

The method was next applied to the observed data. 
With the SDSS sample, we searched for the signatures of the five types of nontranslational isometries, i.e., screw motions of twists $\theta_{\mathrm{twist}}=\pi, 2\pi/3, \pi/2$, and $\pi/3$, and glide reflection, whose results are shown in Figure \ref{ResultData} as was done in Figure \ref{ResultMock}.
It is obvious that, for every type of isometry, the observed data generally follow the prediction of the simply connected space model. 
There are no such robust spikes as seen in Figure \ref{ResultMock}; however, we notice some spikes that could possibly be produced by ghost images. 
In this paper, we consider the spikes that satisfy both $n_{\mathrm{max}}\geq 4$ and $r \geq 1.5$ to be candidates for the topological signatures. 
This criterion is somewhat arbitrary; we rather aim to demonstrate our methodology here.

\begin{figure}
\includegraphics[trim=10 5 5 5,width=9.5cm]{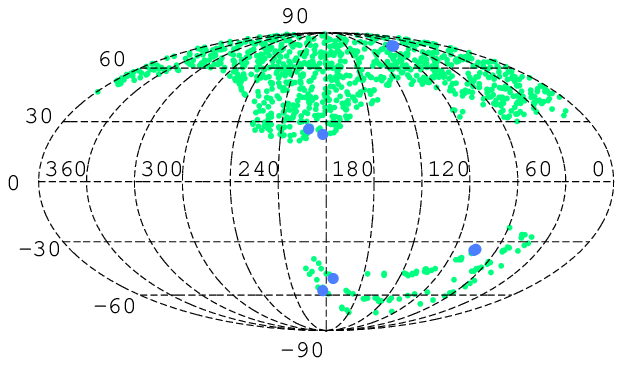} 
\caption{Candidates for the ghost images in the SDSS sample. 
Blue points indicate a pair of isometric quadruples linked by a left-handed sixth-turn screw motion. 
Details of the candidate are summarized in Table \ref{CandidatesTable}.
}
\label{SkyMapCandidates}
\end{figure} 

For each candidate selected from the criterion, we identified the quasars that are responsible for the spike, and then excluded the false detections resulting from ``multi-counts''. 
For example, suppose that there is a pair of twins $(\Vec x_i, \Vec x_j)$ and $(\Vec x_k, \Vec x_l)$ that are both linked by the same isometry $\gamma$ to the same twin $(\Vec x_m, \Vec x_n)$, i.e., $\Vec x_m \approx \gamma \Vec x_i \approx \gamma \Vec x_k$ and $\Vec x_n \approx \gamma \Vec x_j \approx \gamma \Vec x_l$. 
Then the score of $\gamma$ will be overestimated since $(\Vec x_i, \Vec x_j)$ and $(\Vec x_k, \Vec x_l)$ are practically the same images but are double-counted. 

We excluded such cases by eye to leave one candidate for the left-handed sixth-turn screw motion, whose parameters are $n=(-0.28,-0.82,0.50), \ (X,Y)=(1.9,11.2) \ \mathrm{Gpc}$, and $L=4.2 \ \mathrm{Gpc}$ (see Table \ref{CandidatesTable}). We find a pair of isometric  quadruples that are linked by this isometry whose celestial distribution is displayed in Figure \ref{SkyMapCandidates}. In Figure \ref{ResultData}, the corresponding spikes, whose heights are $n_{\mathrm{max}}=4.53$ and $r=1.78$, are marked with vertical arrows. As seen in the figure, the candidate is not statistically significant since we can find several spikes with comparable heights in the simply connected simulations. If the candidate is real, there should be some ghost images (discrete sources or CMB circles) caused by translation of length $6L \approx 25 \ \mathrm{Gpc}$, but it would be hard to observe them since the deduced direction of screw axis corresponds to low Galactic latitude, $b \approx -18^{\circ}$.

Until now, we have only considered 3D positions of objects to identify (the candidates of) ghost images. 
We mention here some possible additional tests based on their astrophysical properties.  
First, it is confirmed from observations of nearby AGNs that their optical continuum emission does not change color \citep{2010ApJ...711..461S} despite their variable flux, so that the ultraviolet--optical spectral index might be used to distinguish real ghosts from false positives. 
The mass of the central black hole ($M_{\mathrm{BH}}$) would also be useful: For two quasars to be the ghosts of each other, they must satisfy $M_{\mathrm{BH, low}z} \geq M_{\mathrm{BH, high}z}$, since $M_{\mathrm{BH}}$ should monotonically increase with time. 
However, it is unknown whether the optical color remains constant over a long period of time ($\lesssim 100 \ \mathrm{Myr}$), and at present estimation of $M_{\mathrm{BH}}$ has large uncertainties. 
Therefore, these additional tests are currently difficult to perform because of both the observational uncertainties and our ignorance of the nature of AGNs, but they may be possible in the future. 
Cross-checks from independent tests, which are based on other catalogs of objects or CMB maps, are more realistic at present. 
For example, \cite{2013arXiv1302.4425R} discussed a further test on the 3-torus model that was identified from the CMB data analyses \citep{2008CQGra..25v5017A}, which makes use of bright star-forming galaxies at redshifts $z\lesssim 6$.

\begin{figure}
\includegraphics[trim=0 0 0 0,width=8.5cm]{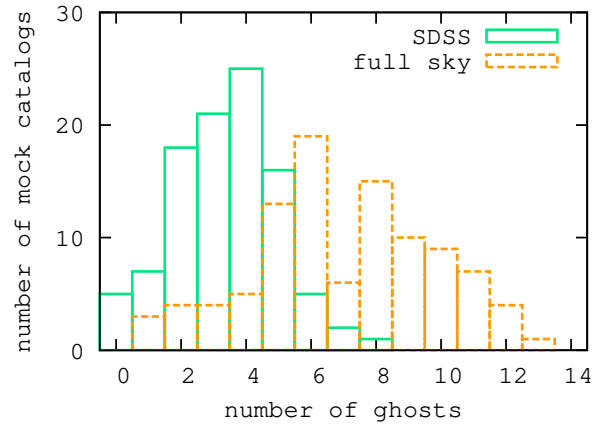} 
\caption{The distribution of the number of ghosts in 100 mock catalogs for two cases of sky coverage. 
{\it Green histogram:} The SDSS DR7 sky coverage. 
{\it Orange histogram:} Full sky ($|b|\geq 30^{\circ}$) coverage. 
If the full-sky data were available, the detectability of topological lensings based on this test would be enhanced by about a factor of 2. 
}
\label{GhostHisto}
\end{figure}

We note that our results do not exclude models whose signatures were not found: In other words, even if the sixth-turn screw motion candidate were rejected by other tests, there would still remain the possibility that we live in a ``small'' multiply connected Universe. 
For quantifying it, we generated 100 mock catalogs in multiply connected space (with an analysis of one catalog being shown in Figure \ref{ResultMock}). 
Among them, only 24 catalogs contained $\geq 5$ ghosts, and 49 contained $\geq 4$ ghosts (Figure \ref{GhostHisto}). 
Therefore, even if real space has the same topology as assumed here, the possibility of detecting the topology is less than $50 \%$ (since by the choice of $s_{\mathrm{min}}=3$ in Equation (\ref{StepFunc}), we need $\geq 4$ ghosts for the detection).
 
This somewhat low detectability is due to the cosmic variance and to the patchy sky coverage of SDSS in the South Galactic cap (see Figure \ref{SkyMap}): Ghost images must be located on the narrow stripe regions or we cannot observe them. 
To robustly constrain the space topology, we need the spectra of quasars over a wider and continuous region in the South Galactic cap.  
As an example, mock catalogs with ``full sky'' coverage except for the low-Galactic latitudes, i.e., $|b| \geq 30^{\circ}$, were generated in the multiply connected space. 
Of the 100 realizations, 84 contained $\geq 5$ ghosts and 89 contained $\geq 4$ ghosts (Figure \ref{GhostHisto}). 
This explicitly shows that the detectability of this test can be much improved by additional data in the Southern sky. 
Ongoing and future wide-field surveys, e.g., SDSS-III/BOSS, VST-ATLAS, LSST, and  Euclid, are expected to provide such data.


\begin{deluxetable*}{ccccccccc}
\tablewidth{17cm} 
\tablecaption{Candidates for the topologically lensed quasars} 
\tablehead{
\colhead{} & \multicolumn{2}{c}{{\small Original image\tablenotemark{a}}} & \colhead{} & \colhead{} & \colhead{} & \multicolumn{2}{c}{{\small Ghost image}} & \colhead{} \\
\colhead{SDSS name} & \colhead{RA [deg]} & \colhead{Dec [deg]} & \colhead{redshift} &
\colhead{} &
\colhead{SDSS name} & \colhead{RA [deg]} & \colhead{Dec [deg]} & \colhead{redshift}
}
\startdata 
\\[-0.2cm]
\multicolumn{9}{c}{{\small Sixth-turn screw motion}} \\
\multicolumn{9}{c}{{\scriptsize left-handed, $\Vec n=(-0.281,-0.817,0.503), \ L=4.2 \ \mathrm{Gpc}, \ (X,Y)=(1.9,11.2) \ \mathrm{Gpc}$}} \\ \cline{2-7} \\
 $024447.79-081606.0$ & 41.199138 & -8.268347 & 4.0678 
 & \multirow{4}*{{\small $\longmapsto$}} & 
 $075612.88+291844.8$ & 119.053681 & 29.312464 & 4.2358 \\

 $025204.28+003137.0$ & 43.017861 & 0.526970 & 4.1490 
 & &
 $073146.99+364346.4$ & 112.945826 & 36.729580 & 4.0181 \\

 $221320.00+134832.5$ & 333.333352 & 13.809046 & 4.1267
 &  &
 $134225.36+363815.1$ & 205.605703 & 36.637536 & 4.0535 \\

 $221705.72+135352.7$ & 334.273851 & 13.897980 & 4.3412 
 &  &
 $134009.78+365813.2$ & 205.040760 & 36.970344 & 4.1928 \\[-0.2cm]
\enddata 
\label{CandidatesTable}
\tablenotetext{a}{In this table, quasar images are called originals if they have positive $x$ coordinates (Equation \ref{Coordinates}), i.e., $0^{\circ} \leq \alpha \leq 90^{\circ}, \ 270^{\circ} \leq \alpha \leq 360^{\circ}$; otherwise, they are called ghosts.}
\tablerefs{\cite{2010AJ....139.2360S}}
\end{deluxetable*}

\acknowledgements{
We thank the anonymous referee for the careful reading of this paper and the useful comments that improved the quality of the paper.
We thank Takeo Minezaki and Mitsuru Kokubo for the useful discussions and suggestions. 
HF thanks Ralph Aurich and Sven Lustig for precious suggestions on the methodology and Jeffrey Weeks for helpful comments on the property of 3D flat manifolds. 
HF is also thankful for the fellowship received from the Japan Society for the Promotion of Science (JSPS).
}

\bibliography{apj-jour,Reference}
\end{document}